\documentstyle[aps,pre,multicol]{revtex}

\begin{document}
\draft
\title{Studies of Chaotic Dynamics in a Three-Dimensional\\ 
 Superconducting Microwave Billiard}

\author{H. Alt$^{1}$, 
        H.-D. Gr\"af$^{1}$, 
        R. Hofferbert$^{1}$,
        C. Rangacharyulu$^{2}$, 
        H. Rehfeld$^{1}$, 
        A. Richter$^{1}$,\\ 
        P. Schardt$^{1,}$\footnote{Present address: 
                         Siemens AG, Bereich Medizinische Technik,
                         D-91052 Erlangen, Germany} and 
        A. Wirzba$^{1}$}
\address {$^{1}$
   Institut f\"ur Kernphysik, Technische Hochschule Darmstadt,\\
   D-64289 Darmstadt, Germany\\
$^{2}$
    Department of Physics, University of Saskatchewan,\\
   S7N OWO, Canada
}
\date{\today}
\maketitle
\begin{abstract}
We present first measurements on a superconducting
{\it three-di\-men\-sio\-nal}, partly chaotic microwave billiard shaped
like a small deformed cup. 
We analyze the statistical properties of the measured spectrum
in terms of several methods originally derived for quantum
systems like eigenvalue statistics and periodic orbits and obtain
according to a model of Berry and Robnik a mixing parameter 
of about $25\%$. In numerical simulations
of the classical motion in the cup 
the degree of chaoticity has been estimated. 
This leads to an invariant chaotic Liouville measure
of about $45\%$. The difference between this figure and the mixing
parameter is due to the 
limited accuracy of the statistical analysis, caused by both,
the fairly small number of 286 resonances
and the rather poor desymmetrization
of the microwave cavity.  
Concerning the periodic orbits of the classical system we present
a comparison with the length spectrum of the resonator and introduce
a new bouncing ball formula for electromagnetic billiards.
\end{abstract}
\pacs{PACS number(s): 05.45.+b, 41.20.Bt, 84.40.Cb}
\begin{multicols}{2}
\noindent
\narrowtext

\section{Introduction}

\label{sec1}
In the last few decades the theoretical investigation of two-dimensional
Euclidian and Riemannian geometries, so-called billiards, has led to a very 
fruitful new discipline in non-linear physics 
\cite {Birkhoff,Bunimovich,Sinai}.
Due to the conserved energy of the ideal particle propagating inside 
the billiard's
boundaries with specular reflections 
on the walls, the plain billiard belongs to
the class of Hamiltonian systems with the lowest degree of freedom in
which chaos can occur and this does only depend on the given boundary shape.
Because of their simplicity two-dimensional billiards are in particular
adequate to study the behaviour of the particle in the corresponding 
quantum regime \cite {McDKau79,LesHouches89,Steiner}
where spectral properties are completely described by the stationary
Schr\"odinger-equation
\begin{equation}
H\Psi(\vec{r})=-\frac{\hbar^2}{2m}\Delta\Psi(\vec{r})=E\Psi(\vec{r})
\end{equation}
inside the domain $\cal{G}$ 
with Dirichlet boundary conditions on the walls
\begin{equation}
\Psi(\vec{r}) |_{\partial\cal{G}}=0.
\end{equation}
In this context the investigation of "Quantum Chaos" has become one of the 
most fascinating
goals of theoretical physics at the end of this century 
\cite{Gutzwiller,BerryBak}.

About five years ago experimentalists have even found very effective
techniques to simulate the quantum billiard
problem with the help of macroscopic
devices. Due to the equi\-va\-lence of the stationary
Schr\"odinger-equation and the classical Helmholtz-equation in two dimensions
one is able to model the billiard by a similarly shaped electromagnetic
cavity \cite {SS90a,SS92,Sridhar}.
In former publications we have demonstrated the high accuracy of large
ensembles of measured eigenvalues as well as of 
resonance shapes and the attached
widths in {\it superconducting} cavities formed like desymmetrized Bunimovich
and truncated Hyperbola billiards \cite {Grapa,Brentano,Hyperbel,widths}.

Especially the statistical analysis of measured 
(or numerically simulated) eigenvalue sequences confirms that a
distinction between classical chaotic and regular systems from
the quantum point of view is only possible in the semiclassical
regime (formally spoken for $\hbar\rightarrow 0$) where
the particle's de Broglie wavelength is sensitive to details
of the billiard's borderline.
One of the most surprising and primarily empirical results of these
investigations is the fact that quantum spectra of classical 
chaotic Hamiltonian systems can be described in a universal manner
which only depends on the global symmetry of the underlying dynamics:
The quantum pendants of
time-reversal invariant classically chaotic systems
typically reveal spectral structures which are reproduced 
excellently by statistical properties 
of the Gaussian Orthogonal Ensemble (GOE)
of Random Matrix Theory \cite {Dyson,Mehta,BerryGoe}. On the other
hand classically regular systems usually lead to spectral
fluctuations on the wave dynamical side according to uncorrelated
Poissonlike distributed random numbers \cite{BerryTabor}.

In this article 
we present first investigations which were performed
on a three-dimensional superconducting billiard. Due to
the polarization properties of the electromagnetic fields $\vec{E}$
and $\vec{B}$ inside
the cavity the full vectorial Helmholtz-equations 
\cite{Jackson}
\begin{eqnarray}
(\Delta +\epsilon\mu\frac{\omega^2}{c_0^2})\vec{E}(\vec{r}) & = & \vec{0}\\
(\Delta +\epsilon\mu\frac{\omega^2}{c_0^2})\vec{B}(\vec{r}) & = & \vec{0}
\end{eqnarray}
have to be used with corresponding boundary conditions
\begin{equation}
\vec E_{\|}(\vec{r})|_{\partial {\cal G}}=\vec{0}~~~\mbox{and}~~~
\vec B_{\perp}(\vec{r})|_{\partial {\cal G}}=\vec{0} \label {Rand}
\end{equation}
on the walls which are assumed to be ideally conducting. Of course
the analogy with the corresponding scalar Schr\"odinger-equation
in the same geometry is fully lost. 
Instead of talking of the semiclassical limit we have to describe
the classical electromagnetic billiard in this region in terms of ray-optical
characteristics, where features of the periodic orbits inside the geometry
dominate the corresponding wave optical side.

Three-dimensional systems have so far only scarcely been investigated
experimentally. To our knowledge first experiments with electromagnetic
waves in cavities simulating acoustic wave phenomena in rooms were
performed by Schr\"oder \cite {Schroeder}. Acoustic model statistics
in metal blocks have also been investigated \cite {Weaver,Guhr}. Very
recently statistical properties of eigenfrequency distributions in
asymmetrically shaped microwave cavities have been reported
\cite {Deus}.
Theoretically quantum effects as well as their electromagnetic counterparts  
in three-dimensional systems were treated
in Refs. \cite{BalBloch,Primack,Eckardt}.

\section {Experiment}

\label {sec2}
We have investigated a small deformed 3D-cavity made from an open deep drawn
Niobium cup 
(a gift, which we had received from the director of the Continuous
Electron Beam Accelerator Facility, CEBAF, Hermann Grunder)
with a welded lid to cover it. Before welding the two parts together,
the cup was deformed at its open end to destroy its rotational symmetry.
Its measures are given in 
Fig.1. The total volume of the resonator was determined by filling
it with water and measuring the additional weight, yielding a
volume of $(122.5 \pm 1.9)\mbox{cm}^3$. 
The shape of the
cup can be approximated by cutting a three-axial ellipsoid
twice perpendicular to its main axis (see also Fig.1) 
and then deforming it slightly. The microwaves were transmitted into and
out of the cavity by two small antennas in the lid. The whole resonator
was cooled down in Helium-atmosphere 
at a temperature of 2K and a
pressure of 38mbar in one of the cryostats of the superconducting
Darmstadt electron linear accelerator S-DALINAC \cite {Beschleuniger}
together with the accelerating
structures. With this setup we have measured microwave spectra in
transmission (antenna 1 for input and antenna 2 for output)
as well as in reflection
(same antenna for excitation and detection) using a 
Hewlett Packard network analyzer (model HP8510B) in a frequency range
between 0 and 20GHz, respectively. 

Spectra were taken in 10kHz steps and  
Fig.2 shows an extraction of the transmission spectrum between 15 and
20GHz where typical Q-values of up to $10^5$ 
and signal-to-noise ratios of up to 50dB
were obtained. By comparing the three measured spectra 286 resonances
could be consistently identified which form the base of all following 
investigations.

For an analysis of this set however one has to take into 
account the following point:
Since the geometry of the cup is very close to one which
possesses two symmetry planes ($xz-$ and $yz-$
plane, see Fig.1) the given set of resonances 
in the analysis is always compared to 
a superposition of four independent subspectra. 

\section {Results and discussion}

\label {sec3}

\subsection {Density of eigenmodes}

In order to derive meaningful statistical measures for 
the given eigenvalue sequence
it is first necessary to extract the smooth part of the resonator's 
density of eigenmodes which is given by the generalized electromagnetic
Weyl-formula \cite{Weyl1,Weyl2,BalHilf,BalDup}
\begin{equation}
\label{density}
\rho^{smooth}(f)=\frac{8\pi}{c_0^3}|{\cal{G}}|f^2+const.~~,
\end{equation}
where $|{\cal{G}}|$ denotes the volume of the cavity and
$f$ the upper frequency limit of the given spectrum.
The constant term contains contributions of the surface's curvature 
as well as of the edges of the cavity.
The total density of eigenmodes contains in addition a fluctuating part
\begin{equation}
\label {densitytotal}
\rho(f)=\rho^{smooth}(f)+\rho^{fluc}(f)=\sum_i\delta(f-f_i)~ ,
\end{equation}
where $f_i$ denotes the eigenfrequencies of the resonances.

It is very instructive to compare Eq.(\ref {density})
with the corresponding expression for the three-dimensional 
Schr\"odinger-problem in the same geometry, i.e.
\begin{equation}
\label{qmdensity}
\rho^{smooth}(f)=\frac{4\pi}{c_0^3}|{\cal{G}}|f^2
-\frac{\pi}{2c_0^2}|{\partial\cal{G}}|f+const.~~.
\end{equation}
Due to the already mentioned
polarization features of the modes, the leading term in the electromagnetic
formula, Eq.(\ref{density}), 
is twice the corresponding term in the scalar problem,
Eq.(\ref{qmdensity}),
because of two transverse directions of polarization relative to the 
axis of propagation. In addition these two polarizations which are
known as TM- and TE-modes in certain geometries
provide linear terms in the smoothed
eigenmode density of the same magnitude
but with different sign. Thus due to cancellation there is 
no linear contribution in Eq.(\ref {density}), 
whereas this term survives in the
scalar case and is proportional to the cavity's surface $|{\partial\cal{G}}|$.
Even if a clear separation in the attached TM- and TE-modes is not possible
in cases of 
arbitrary geometries, like in the case of our cup, this 
linear term vanishes for all piecewise smooth boundaries
\cite {BalHilf,BalDup}.
The constant terms in both Eqs. (\ref {density}) and ({\ref {qmdensity})
are of the same origin, i.e. the curvature 
of the surface and the edge angles of the cavity.

To determine the spectral fluctuations, 
the smooth part of the eigenmode density in
the measured spectrum had to be eliminated. 
For this we constructed from Eq.(\ref {densitytotal})
the staircase function
\begin{eqnarray}
N(f) & = & \int_0^f df'\rho(f')\label{totaltreppe}\\
& = & \sum_{i}\Theta (f-f_i)=
\sum_{i \atop f>f_i}1+\sum_{i \atop f=f_i}\frac{1}{2}\nonumber
\end{eqnarray}
and obtained its fluctuating part
\begin{equation}
N^{fluc}(f)=N(f)-N^{smooth}(f)~ .
\label{treppchen}
\end{equation} 
Since in the case of our billiard
there is no analytical form for the edge contribution of
the constant term of Eq.(\ref {density}) we have fitted a third
order polynomial (without the quadratic term) to the experimental
staircase function, i.e.
\begin {equation}
N^{smooth}(f)=V_1\cdot f^3+V_2\cdot f+V_3~ . \label{Nsmooth}
\end{equation}
In Fig.3 the remaining fluctuating part
of the staircase function can be seen to oscillate around zero as expected,
the fitted constant $V_1$ corresponds to a volume of
$(119.3 \pm 0.7)\mbox{cm}^3$ which is very close to the correct value of 
$(122.5 \pm 1.9)\mbox{cm}^3$. This difference in the leading term
corresponds to an uncertainty of 4 resonances in the measured total
spectrum (neglecting the linear and constant term in Eq.(\ref{Nsmooth})). 

\subsection {Nearest neighbour spacing distribution}

In order to perform a statistical analysis of the given eigenvalue sequence
independently from the special size of the resonator, the spectrum was first
unfolded, i.e. from the measured 
sequence of eigenfrequencies $\left\{f_1, f_2,
\dots, f_i, f_{i+1}, \dots \right\}$ the spacing 
$s_i= \left(f_{i+1}-f_i\right) / \bar{s}$ 
between adjacent eigenmodes was obtained by calculating the 
local average $\bar{s}$ from Eq.(\ref{Nsmooth}). The proper normalization of 
the measured spacings of eigenmodes 
then yielded the desired nearest neighbour
distribution P(s), i.e. the probability for a certain spacing s. 

By comparing P(s) to theoretical expressions one has to take
into account, however, that the mechanically only weakly deformed cup (Fig.1) 
can to a good approximation be constructed of four similar 
quartercups which separately possess the full geometrical information of the
object, in which case the measured spectrum would be
a superposition of 
four independent symmetry classes which are obtained by permuting
the boundary conditions on the cutting planes of the quartercup
from electric to magnetic, respectively. 
Hence one has to be aware of this point in the 
following statistical analysis of eigenmode spacings.

Furthermore, to obtain a quantitative criterion concerning the degree of 
chaoticity in the system the spectrum was analyzed in terms of statistical 
measures from a model of Berry and Robnik \cite{BerRob}
which interpolates between the two limiting cases of pure Poissonian
and pure GOE behaviour for classical regular and chaotic systems,
respectively. The final essence of this model is a mixing-parameter
$q$ which is directly related to the relative chaotic part of the invariant
Liouville measure of the underlying classical phase space in which
the motion takes place. According to the herein embedded
one-to-one connection
between classical phase space and eigenmode density for
the two different regions of regular and chaotic motion, 
a comparison becomes meaningful only for
the highly excited domain of the spectrum
which is not sufficiently covered by the frequency range up to
20GHz investigated here.

In the upper part of
Fig.4 the result for the nearest neighbour spacing distribution $P(s)$
which describes, as already mentioned,
short range correlations between neighbouring unfolded
levels is shown in form of a histogram. 
With respect to the superposition of four independent
symmetry classes, as noted above, the model of Berry and Robnik makes the
following
ansatz for an interpolation between pure Poissonian and pure Gaussian
characteristics
\begin{eqnarray}
P_{4,4}(s,q) & = & e^{-(1-q)s}\bigg( 16(\frac{1-q}{4})^2 erfc^4\big(
{\frac{\sqrt\pi}{2}}{\frac{q}{4}}s\big)+\label {NND44}\\
& &\big[32\frac{(1-q)}{4} \frac{q}{4}
+{\frac{\pi}{2}}{(\frac{q}{4})}^3 4s\big]\times \nonumber\\
& &\exp{(-{\frac{\pi}{4}}{(\frac{q}{4})}^2{s^2})} 
erfc^3\big({\frac{\sqrt\pi}{2}}
{\frac{q}{4}}s\big)+\nonumber\\
& &12(\frac{q}{4})^2 \exp(-2{\frac{\pi}{4}}(\frac{q}{4})^2 s^2) erfc^2\big(
{\frac{\sqrt\pi}{2}}{\frac{q}{4}}s\big)
\bigg)\nonumber~ .
\end{eqnarray} 
For inspection the limiting curves of a superposition of four
independent pure Poissonians ($q=0$), 
which again yields one single Poissonian,
as well as of four independent GOEs ($q=1$) are also represented
in the figure. Note that the superposition results in a clear loss of linear
level repulsion for small spacings in the case of pure GOE statistics.
Furthermore there is not much distinction between Poissonian and GOE 
statistics, anyhow. 

It is obvious that $P(s)$ does not allow to determine the mixing
parameter $q$ with reasonable significance, since
the fluctuations in the data are larger than the
difference between the given smooth curves.
To be free of effects due to the binning of $P(s)$ we have also calculated
the cumulative nearest neighbour spacing distribution
\begin{equation}
I(s)=\int_0^s\,P(s')ds'~,
\end{equation} 
the result is presented in Fig.4, lower part, together with the
curves for the pure distributions, Poisson and 4xGOE.
In this case the measured spacings are very close to
Poissonian behaviour over
a large range of s.
A fit of $I(s)$ to the data yields $q=0.16\pm^{0.25}_{0.16}$, where the
uncertainty was determined from analyzing different subsets
of the measured data.
Because of this rather poor sensitivity
other statistical measures like the number variance and
spectral rigidity had to be applied to the data to determine $q$.

\subsection{Number variance and spectral rigidity}

In order to check for long range 
correlations between the measured levels we
have calculated $\Sigma^2-$ as well as $\Delta_3-$statistics, two
measures originally introduced by Dyson and Mehta \cite{Dyson,Mehta}
for studies in equivalent fluctuations of nuclear spectra.
In this case one is interested in spectral correlations on a scale
which contains L mean level spacings.
Here 
\begin{equation}
\Sigma^2(L):=\big\langle(n(L)-<n(L)>)^2\big\rangle=
<n^2(L)>-L^2
\end{equation}
represents the averaged variance of a number $n(L)$ 
of levels belonging to an interval
of length $L$ on the unfolded axis with mean $<n(L)>=L$.
The quantity $\Delta_3(L)$ is a smoothed and rescaled version
of $\Sigma^2(L)$ and can be calculated from
\begin{equation} \label {Delta3}
\Delta_3(L)=\frac{2}{L^4}\int_0^L dr\,(L^3-2L^2r+r^3)\Sigma^2(r)~ .
\end{equation}
Figure 5 shows the experimental results for these two measures as well 
as the fitted curve 
according to Berry and Robnik \cite{BerRob} deduced from
\begin{equation}
\Sigma^2_{4,4}(L,q) = \Sigma^2_{Poisson}\big((1-q)L\big)+
    4\cdot\Sigma^2_{GOE}\big( q \frac{L}{4}\big)~ , 
\label {Sig44}\\
\end{equation}
where $\Sigma^2_{Poisson}$ and $\Sigma^2_{GOE}$ are the distributions
for the pure cases and one symmetry class. An equivalent formula
is valid for $\Delta_3(L)$. As in the case of the nearest neighbour spacing
distribution the superposition of four pure Poissonians
yields one single Poissonian which can be directly seen from
the analytical expression
\begin{equation}
\Sigma^2_{Poisson}(L)=L~ .
\end{equation}
Inspecting Fig.5
two results have to be noted: Firstly, the experimental
number variance $\Sigma^2$
lies clearly 
between the limiting curves of pure Poissonian and
Gaussian characteristics. A fit of the 
expression (\ref{Sig44}) to the data yields
a mixing-parameter of $\Sigma^2(L)$, $q=0.30\pm^{0.20}_{0.30}$,
which agrees within the given error range with
to the $q$-value derived from
the cumulative nearest neighbour
spacing distribution. On the other hand the  
$\Delta_3(L)-$curve only corresponds within the given
error range of the data with this result, 
i.e. a deviation from pure Poissonian
behaviour is basically not visible.
Secondly, the number variance $\Sigma^2(L)$ clearly
displays saturation above a certain value $L_{max}$.
In fact, Berry showed \cite {Berry400}
that a global and universal semiclassical (or ray-optical) behaviour 
can only be expected between $L_{min}=1$ and the value $L_{max}$
which is determined on one hand by the finite ensemble of resonances,
but above all by the lengths of the shortest periodic orbits of the 
classical system. From Fig.5 we find $L_{max}\approx 10$ for $\Sigma^2$.
Note, that $L_{max}$ of $\Delta_3$ is well approximated by  
four times $L_{max}$ of $\Sigma^2$
(see Eq.(\ref{Delta3}) and \cite {Jost}). Using a modification of 
Berry's expression 
\cite {Berry400} for the electromagnetic case, $L_{max}$ of $\Delta_3$ can be
related to an average length $l_{min}$ 
of the shortest periodic orbits via the
expression 
\begin{equation}
L_{max} = \frac{3c_0}{l_{min}f_{max}}\cdot \frac{N_0}{2}~ ,
\end{equation}
where $N_0$ is equal to the first term $V_1\cdot f^3$ in Eq.(\ref{Nsmooth})
and $f_{max}$ denotes the upper frequency, i.e. $f_{max}=20$GHz. 
The result is $l_{min}\approx0.17$m in fair agreement with a value deduced 
independently in Sect.E below.

In addition to $\Sigma^2(L)$ and $\Delta_3(L)$ we have also calculated 
another long range statistics: the two-level form factor $b_2(t)$
in the corresponding time domain \cite {Jost,Aukolo} which leads
to a so-called ''autocorrelation hole'' in the case of a GOE-like
sequence. It turns out that 
the fluctuations in $b_2$ due to the small number of resonances
are to large for a proper conclusion
and even in the pure 
GOE-case the hole is not very pronounced due to the superposition 
of symmetry classes.

To summarize the statistical investigation, only the cumulative 
nearest neighbour spacing distribution $I(s)$ and the number variance
$\Sigma^2$ lead to a significant deviation from pure Poissonian
behaviour for the given set of resonances. Furthermore the extracted
mixing parameter $q$ represents an upper limit for the chaoticity
since the assumption of a superposition of four symmetry classes 
in the statistical analysis increases the
weight of the pure GOE-contribution with respect to the desymmetrized
case.

\subsection{Classical surface of section}

In order to obtain an independent estimate for the degree of
chaoticity in the system
we have also performed a numerical simulation of classical 
motion inside the 3D-cup. Following an idea of Zaslavsky and Strauss 
\cite {ZasStr} we have approximated the resonator's geometry
by one half of a so-called barrel billiard which can be obtained by cutting
a three-axial ellipsoid in two different 
heights perpendicular to the longest of its axis, see also Fig.1. 
Consequently the surface of the barrel can be described by the curve
\begin{equation}
\frac{x^2}{a^2}+\frac{y^2}{b^2}+\frac{z^2}{c^2}=1
~~~\mbox{with}~~z\in [-h,0]
\end{equation} 
and the resonator was modelled with axes $a=58.5$mm/2,$b=67.6$mm/2
and cuts at $z=0$ and $z=-h=-56.1$mm. With respect to these parameters
the relative geometrical difference 
in the volume between the cup resonator and 
the barrel billiard is about $11\%$.

To characterize the classical motion of a particle inside this
''half of a barrel geometry'' a two-dimensional area preserving
mapping also introduced in \cite {ZasStr} was used for creating
the underlying Poincar\'e surface of section. 
The base of this mapping is simply
a pair of a coordinate and its conjugated momentum $(z,p_z)$ of the particle 
during each reflection on the wall. Note that the momentum
$|\vec{p}|$ is normalized to unity. Figure 6 shows the resulting patterns
in phase space after 16000 collisions with the boundary. Because
the full ellipsoid
is totally regular \cite {Berklas} there exists a class of trajectories
(also shown in Fig.6) manifested by regular stripes on the surface of section
which indicate remaining 
regularities inside the cup. This is due to the fact
that these trajectories do not impact on the plane at
$z=-h$ which exactly is the reason for overfocussing features of the geometry
and consequently the origin of chaos. The relative area of the 
chaotic part of phase space, the so-called ''chaotic sea'', was
estimated by an invariant chaotic Liouville measure of $q\approx45\%$
using a drastically increased number of about $2\cdot 10^5$ wallcollisions
to get a higher accuracy. 
This is bigger than the mixing-parameter 
deduced from $I(s)$ and $\Sigma^2$, but it has to be noted
that the applicability of the Berry-Robnik-model was originally
shown to depend strongly upon reaching 
the semiclassical (or ray-optical) limit
which in the two-dimensional case is established only far beyond thousands
of eigenvalues \cite {Robnik}. The present case of only close to
three hundred eigenvalues at least points to the 
correct tendency for the correspondence
between the classical Liouville measure and the wave dynamical 
mixing-parameter extracted from the data.

\subsection{Periodic orbit theory}

\label {sece}
As a final study of semiclassical features of our wave dynamical system,
we have analyzed the spectrum of classical orbit length $l$ in the 3D-cup
which is directly
related to the measured frequency spectrum by a Fourier transform
of the above mentioned fluctuating part of the eigenmode density
(Eq.(\ref{densitytotal})),
\begin{eqnarray}
\tilde{\rho}^{fluc}(l) & = & \int_{f_{min}}^{f_{max}}\rho^{fluc}(f)
\cdot \exp({i\frac{2\pi}{c_0}lf})\,df\label {Fourint}\\
& = & \int_{f_{min}}^{f_{max}}\big(
\rho(f)-\rho^{smooth}(f)\big )
\cdot \exp({i\frac{2\pi}{c_0}lf})\,df \nonumber~.
\end{eqnarray}  
Here $f_{min}$ and $f_{max}$ denote the borders of the measured frequency
range, i.e. 0 and 20GHz. Figure 7 shows in the upper part the value 
$|\tilde{\rho}^{fluc}(l)|^2$
in a range of orbit lengths $l$ up to $0.5$m, and in the lower part 
some periodic orbits
obtained from numerical simulations on the barrel billiard are presented.
As can be seen from the arrows below the abscissa several of those
orbits correspond well to the locations of peaks in the
Fourier spectrum of the data. The shortest periodic orbit which 
bounces between the bottom and the lid has a length of $l=0.1122$m,
which is close to the independent estimate of Sect.C above.
It is the so-called shortest bouncing ball orbit in the cup.

\subsection{3D-bouncing ball orbit}

This bouncing ball orbit has been analyzed in a
more quantitative manner. Following \cite {Smilansky} we have 
calculated an additional term of the smooth part of the staircase function,
Eq.(\ref{Nsmooth}),
attached to the three-dimensional bouncing ball orbit propagating
periodically between the bottom and the lid of the cup. 
Again the contributions for both polarization classes have to be
considered separately. The result for the sum of both parts
as pointed out in the Appendix
is given by
\begin{eqnarray}
N^{bbo}_{em}(X) & = & N^{bbo}_{TM}+N^{bbo}_{TE}\label {3DBB2}\\
& = & \frac{\pi S}{2 h^2}\bigg(\sum_{0 < n < X}(X^2-n^2)-
\frac{2}{3}X^3+\frac{1}{2}X^2\bigg)\nonumber~,
\end{eqnarray}
where $X=kh/\pi=2hf/c_0$ and $h=56.1$mm, the half
of the bouncing ball orbit's length, i.e. the heigth of the cup. 
The parameter $S$ denotes the area on which
this orbit exists, i.e. the size of the cup's bottom.
The upper part of Fig.7 also shows the result
for the remaining length spectrum after 
extracting this contribution. In fact,
the peak belonging to the correct length of the bouncing ball orbit
does not vanish completely because it is due to 
a superposition of two adjacent periodic orbits, the bouncing ball orbit
at $l=0.1122$m and a stable periodic orbit at $l=0.1170$m, as seen from the
lower part of Fig.7.

In order to verify the new electromagnetic bouncing ball formula, 
Eq.(\ref {3DBB2}), 
independently from this result
we also have tested it using a set of $\approx20000$
eigenmodes of a regular box. This system is classically integrable,
thus it is possible to calculate the electromagnetic eigenfrequencies
analytically. The box also allows to study the different polarizations,
TE- and TM-modes, in more detail and especially their systematic
degeneration which directly follows from the analytical expression for
the eigenfrequencies
\begin{eqnarray}
& & f_{u,v,w} = \frac{c_0}{2}\sqrt{(\frac{u}{A})^2+
(\frac{v}{B})^2+(\frac{w}{C})^2}~~\mbox{with} \\
\mbox{TM} & : & ~u,v=1,2,3,...~~w=0,1,2,...\nonumber\\
\mbox{TE} & : & ~u,v=0(\mbox{either}~u~\mbox{or}~v),1,2,...~~w=1,2,3,...~~.
\nonumber
\end{eqnarray}
In order to avoid further accidental number theoretical degeneracies  
the lengths of the three edges A, B and C were chosen as
$A=0.2$m, $B=\gamma A$ and $C=\gamma^2 A$, where $1/\gamma=(\sqrt{5}-1)/2$,
the ratio of the golden mean. 
Following Eq.(\ref{density}) we have calculated
the exact smooth part of the staircase function for the given set of
eigenfrequencies. Here the linear term as well as the constant term
are given analytically without any free parameter. Extracting this smooth
part from the total staircase of the box, Eq.(\ref {totaltreppe}), 
we obtained the fluctuating part, which is shown in the upper part of Fig.8.
The present choice of the edges A, B and C yields a very clear oscillation
in this fluctuating part, which is dominated by one certain classical
bouncing ball orbit (also given in Fig.8).
The length of this orbit, $l_{bbo}$, 
corresponds to the period of the observed
oscillation $\Delta f=c_0/l_{bbo}=c_0/{2A}\approx 750$MHz.
Using the area $S_{bbo}=B\cdot C$ for this certain orbit, we have calculated
the contribution which follows from Eq.(\ref{3DBB2}). 
This curve is also
represented in the upper part of Fig.8 and reproduces the data very well
as it can be seen in the lower part of the same figure where the 
bouncing ball contribution has been extracted from the staircase function,
i.e. where the modulation of $\Delta f$ is not present anymore.
Note
that the impressive amplitude of the given modulation is due to the
ratio $2\pi\cdot S_{bbo}/l_{bbo}^2=\pi/2\cdot \gamma^3\approx 6.65$ 
in Eq.(\ref{3DBB2})
which is the largest contribution in comparison to the other
possible bouncing ball orbits and also much larger than the 
corresponding amplitude for the 3D-cup, 
$\pi S/(2 h^2) \approx 0.45$. 
 
To perform a more obvious test of Eq.(\ref {3DBB2}) we again have calculated
the Fourier spectrum, Eq.(\ref{Fourint}), 
of the calculated set of eigenfrequencies. The result is shown in Fig.9. 
The bouncing ball
orbit leads to an impressive peak at a length $l_{bbo}=0.4$m.
Using Eq.(\ref {3DBB2}) in the same way, i.e.
calculating the Fourier transformed of $N^{bbo}_{em}$ we are able to extract
this peak without any remnant as well as contributions of its multiples,
which indicates that the orbit has been described
correctly by Eq.(\ref {3DBB2}). Because of this result,
which was achieved successfully for several boxes of different proportions,  
we are certain that the remaining peak in the length spectrum of the 3D-cup
in Fig.7 is due to the existence of the stable periodic orbit
of 0.1170m length. 

As an additional test of statistical measures, we have calculated
the nearest neighbour spacing distribution, the $\Sigma^2-$ and the
$\Delta_3-$statistics for both polarizations of the 
box separately. This yields pure Poissonian characteristics accross
the whole range of spacings as expected for the given regular system.
As an example, the result for the nearest neighbour spacing distribution
using the first 9978 TE-modes is presented
in Fig.10.

Return to the 3D-cup:
For completion we have calculated the statistical measures
of the short and long range correlations 
after extraction of the bouncing ball contribution,
i.e. we have reunfolded the experimental 
spectrum in adding Eq.(\ref {3DBB2}) to the
standard Weylian, Eq.(\ref {Nsmooth}), and repeated our statistical
analysis. The effect is rather weak, as can be expected because
of the small
amplitude of Eq.(\ref {3DBB2}) in the case of the cup. 
The procedure leads to a very
slight correction
of the $\Sigma^2-$ and the $\Delta_3-$curves towards the pure GOE-limit
but its influence on the mixing parameter is much smaller than
the quoted uncertainties.

\section{Conclusions}

In summary we have applied several original quantum methods like 
eigenvalue statistics and periodic orbits in the given case of
a purely classical, vectorial wave phenomenon.  
Although the spectrum was checked for completeness using the correct
electromagnetic Weyl-formula, Eq.(\ref {Nsmooth}), the fluctuations 
around this mean behaviour are not very sensitive for the details 
of the boundary in the present case of only 286 resonances. 
On the other
hand the system was analyzed in terms of periodic orbit theory.
As the length spectrum of the resonator leads to more detailed
conclusions (estimations for the saturation of $\Sigma^2$ and
$\Delta_3$ via the shortest periodic orbits, identification of classicals
orbit lengths and extraction of the first bouncing ball orbit) this
analysis reveals as a helpful tool for investigations in
the near semiclassical limit.
Comparing the classical degree of chaoticity 
($q_{cl}\approx 0.45$) with the corresponding mixing parameter deduced
from spectral statistics according to the model of Berry and Robnik,
only the cumulative nearest neighbour spacing distribution $I(s)$ and
the number variance $\Sigma^2$ significantly deviate
from the pure Poissonian behaviour and show the correct tendency
($q>0$). 
The uncertainties of the analysis are not due to fundamental difficulties
but rather a consequence of the poor desymmetrization and the small
size of the cup, yielding only 286 resonances in the accessible
frequency range. 
Therefore, in order to study the ''three-dimensional
Helmholtz-Chaos'' in a more quantitative and proper way we have already
started measurements on a precisely
manufactured, fully desymmetrized
3D-Sinai-Billiard realized by 1/48 of a cube with a centered sphere
\cite {Primack}.
Besides testing the usual statistical measures we will especially
investigate the ray-optical generalization of
Gutzwiller's quantum trace formula following the very advanced
ideas of Balian and Duplantier \cite {BalDup}.

\begin{acknowledgements}
We like to thank F. Neumeyer and S. Strauch
for helpful advice concerning the 
numerical calculations
and our mechanical workshop for
the excellent fabrication of the Niobium resonator. 
One of us, A.R., is grateful to H. Grunder for the present of the
CEBAF cup. This work has been supported
by the Sonderforschungsbereich 185 ''Nichtlineare Dynamik'' of the Deutsche
Forschungsgemeinschaft, in part 
by the Bundesministerium f\"ur Bildung und
Forschung under contract number 06DA665I, and through 
a Max-Planck-Forschungspreis.
\end{acknowledgements}

\appendix

\section*{The bouncing ball staircase function in $D$ dimensions} 

First, we construct the general contribution of a bouncing ball 
orbit in a  $D$-dimensional cavity ($D$ $>$ 1)  
to the spectral staircase function of non-relativistic quantum 
mechanics, $N_D^{bbo}(k)$ with $k=2\pi f/c_0$,  
and secondly we apply these general results 
to the electromagnetic case in $D=3$
dimensions.
As usual the spectral staircase function is given as the integral over
the spectral density, $\rho_D^{bbo} (E)$,
\begin{eqnarray}
   N_D^{bbo}(k) & = & \int_{0}^{k} d \tilde k\, \rho_D^{bbo}(\tilde k)
                \label{N_k}\\ 
                & = & \int_{0}^{k} d \tilde k\, \frac{d \tilde E} {d \tilde k}
                      \rho_D^{bbo} (\tilde E) 
                = \int_{0}^{k} d \tilde k\, \frac{\hbar^2 \tilde k}{m}
                      \rho_D^{bbo} ({\tilde E}) \nonumber~ ,
\end{eqnarray}
where $m$ is the mass of the non-relativistic particle inside the cavity.
The spectral density in turn is given by the imaginary part of the trace of
the non-relativistic Green's function inside the cavity:
\begin{eqnarray}
 \rho_D^{bbo} (E) &=& -\lim_{\epsilon\to 0}
               \frac{1}{\pi} {\rm Im} Tr\big( G(E+i\epsilon)
             \big)       \label{d_E}\\
               &=& -\lim_{\epsilon\to 0}\frac{1}{\pi} {\rm Im} 
\sum_{n {> \atop {(-)}} 0} \frac{S_{\perp}^{(D-1)}}{(2\pi)^{D-1}}
         \times \nonumber\\
              & & \int d^{D-1} k_\perp \,
       \frac{1}{E+i\epsilon - \frac{ \hbar^2}{2m}
     \left ( {k}_\perp^2 +\frac{n^2 \pi^2}{l^2} \right )} \nonumber~ .
\end{eqnarray}
Here $l$ is half of the total length of 
the bouncing ball orbit and $S_\perp^{(D-1)}$ 
is the size of the smallest of the two parallel  
$\mbox{$D-1$}$ dimensional ``surfaces'' between
which the orbit bounces. The integer $n$ labels the 
modes  along the bouncing ball orbit with 
$n > 0$ or $n\geq 0$ for Dirichlet or Neumann boundary 
conditions on the
two surfaces, respectively. The  momentum component along the orbit 
is $n\pi/l$. 
The $D-1$ dimensional momentum integration  perpendicular to the 
bouncing ball orbit is the free unrestricted one. 
The factors in front of the integral are 
the phase space normalization factors in $\mbox{$D-1$}$ dimensions.
After the insertion of Eq.(\ref{d_E}) into Eq.(\ref{N_k}), the spectral 
staircase function reads
\begin{eqnarray}
 N_D^{bbo}(k) & = & \int_0^{k^2} d(\tilde k^2)\, 
\frac{S^{(D-1)}_\perp}{(2\pi)^{D-1}} 
 \sum_{n {> \atop {(-)}} 0} \int 
 d\Omega_{k_\perp}^{(D-1)}\times       \label{N_k_2}\\
& & \int_0^{\infty} dk_\perp\, k_\perp^{D-2} 
    \delta \left(\mbox{$k^2_\perp +\frac{n^2\pi^2}{l^2} 
     -{ \tilde k}^2 $} \right )\nonumber
 ~ ,
\end{eqnarray}
where $\int d\Omega_{k_\perp}^{(D-1)} = 2 \pi^{(D-1)/2}/\Gamma((D-1)/2)$ 
is the  angular integral  in $\mbox{$D-1$}$ dimensions. In
writing down Eq.(\ref{N_k_2}) we used the following 
relation:
\mbox{$ \lim_{\epsilon\to 0} {\rm Im} (x-x_0-i\epsilon)/\pi= \delta(x-x_0)$}.
Now the integrations in Eq.(\ref{N_k_2}) become simple 
because of the delta function.
The final result for the bouncing ball orbit of a non-relativisitic particle
inside a cavity in $D>1$ dimensions is therefore
\begin{eqnarray}
 N_D^{bbo}(k) & = & \frac{S_\perp^{(D-1)}}
{\Gamma( \frac{D-1}{2} ) (2\sqrt{\pi})^{D-1}}\times \,\\
& & \frac{1}{\frac{D-3}{2}+1} \sum_{0 {< \atop {(-)}} n < \frac{kl}{\pi}}
 \left ( k^2 - \frac{n^2 \pi^2}{l^2} \right )^{(D-3)/2+1}\nonumber~ .
\end{eqnarray} 
This formula applied to two dimensions reads
\begin{eqnarray}
 N_2^{bbo}(k)= \frac{L}{\pi}   \sum_{0 {< \atop {(-)}}  
  n < \frac{kl}{\pi}}
 \sqrt{ k^2 - \frac{n^2 \pi^2}{l^2}} \ ,
\end{eqnarray}
where $L=S_\perp^{(1)}$ is the length 
perpendicular to the bouncing ball orbit.
It agrees of course with the result 
of Ref.\cite{Smilansky}. The three-dimensional 
expression
has the form
\begin{equation}
 N_3^{bbo}(k)= \frac{S}{4\pi}   \sum_{0 {< \atop {(-)}} 
   n < \frac{kl}{\pi}}
 \left( k^2 - \frac{n^2 \pi^2}{l^2} \right ) ~,
 \label{N_k_3D}
\end{equation}
where $S=S_\perp^{(2)}$ denotes the plane perpendicular to the bouncing
ball orbit.
The bouncing ball contribution in three 
dimensions to the {\em fluctuating} part
of the spectral staircase function is now given by  formula (\ref{N_k_3D})
minus its contribution to
the Weyl or smooth part of the spectral staircase function which has to be 
subtracted in order to avoid 
double counting. The smooth part follows from Eq.(\ref{N_k_3D})
via the application of the Euler-MacLaurin formula 
$\frac{1}{2} F(0) + F(1) + F(2) + \cdots F(N-1) +\frac{1}{2} F(N) =
\int_0^N dx\, F(x) +\dots$, 
where \mbox{$F(x)\propto (k^2-x^2\pi^2/l^2)$} and 
the dots correspond to the fluctuating contribution here.
Thus we have
\begin{eqnarray}
 N_3^{bbo,{\rm smooth}}(k) &=& \frac{S}{4\pi}\left( 
 \int_0^{kl/\pi} d x \, \left( k^2 -\frac{x^2\pi^2}{l^2}
\right) \mbox{\raisebox{-.8ex}{$\,{\stackrel{\mbox{$-$}}{(+)}}\,$}} 
     \frac{1}{2} k^2 \right )\nonumber \\
     &=& \frac{S l k^3}{6 \pi^2} \mbox{\raisebox{-.8ex}{$\,
         {\stackrel{\mbox{$-$}}{(+)}}\,$}} \frac{S k^2}{8\pi} \ ,
\end{eqnarray}
where the last term takes into account the factor $1/2$ in front of the 
first term, $F(0)$, in the Euler-MacLaurin formula and the extra mode in the
Neumann case. We do not get an extra contribution from the upper boundary, as 
$F(N)$  vanishes in the average
in our case. In summary, in non-relativistic quantum mechanics, the
bouncing ball contribution to the fluctuating part of the spectral staircase
function in a three-dimensional cavity reads 
\begin{equation}
 N_3^{bbo,{\rm fluc}}(k) = 
\frac{S}{4\pi}   \sum_{0 {< \atop {(-)}} n < \frac{kl}{\pi}}
 \big( k^2 - \frac{n^2 \pi^2}{l^2} \big ) - \frac{S l k^3}{6 \pi^2} 
 \mbox{\raisebox{-.8ex}{$\,{\stackrel{+}{(-)}}\,$}}  \frac{S k^2}{8\pi}~, 
 \label{N_k_3D_fluct}
\end{equation}
where the upper signs/inequalities apply to 
Dirichlet boundary conditions on the two surfaces
between which the orbit bounces, whereas the lower signs/inequalities 
refer to the Neumann case.
In the electromagnetic case these 
two contributions correspond to the magnetic
and electric bouncing ball modes, respectively, see Eq.(\ref{Rand}), which
decouple for modes along the bouncing ball orbit. 
Therefore we can just sum both
terms to get the final expression for the bouncing ball contribution to
the fluctuating part of the staircase 
function for an electromagnetic cavity in
three dimensions: 
\begin{equation}
{N}_{em}^{bbo,{\rm fluc}} (X) = \frac{\pi S}{2 l^2}\bigg(  
  \sum_{0 < n < X}
 \big(X^2 - n^2 \big) - \frac{2}{3} X^3 +\frac{1}{2} X^2 \bigg) \ 
\end{equation}
with $X=kl/\pi$.

\clearpage
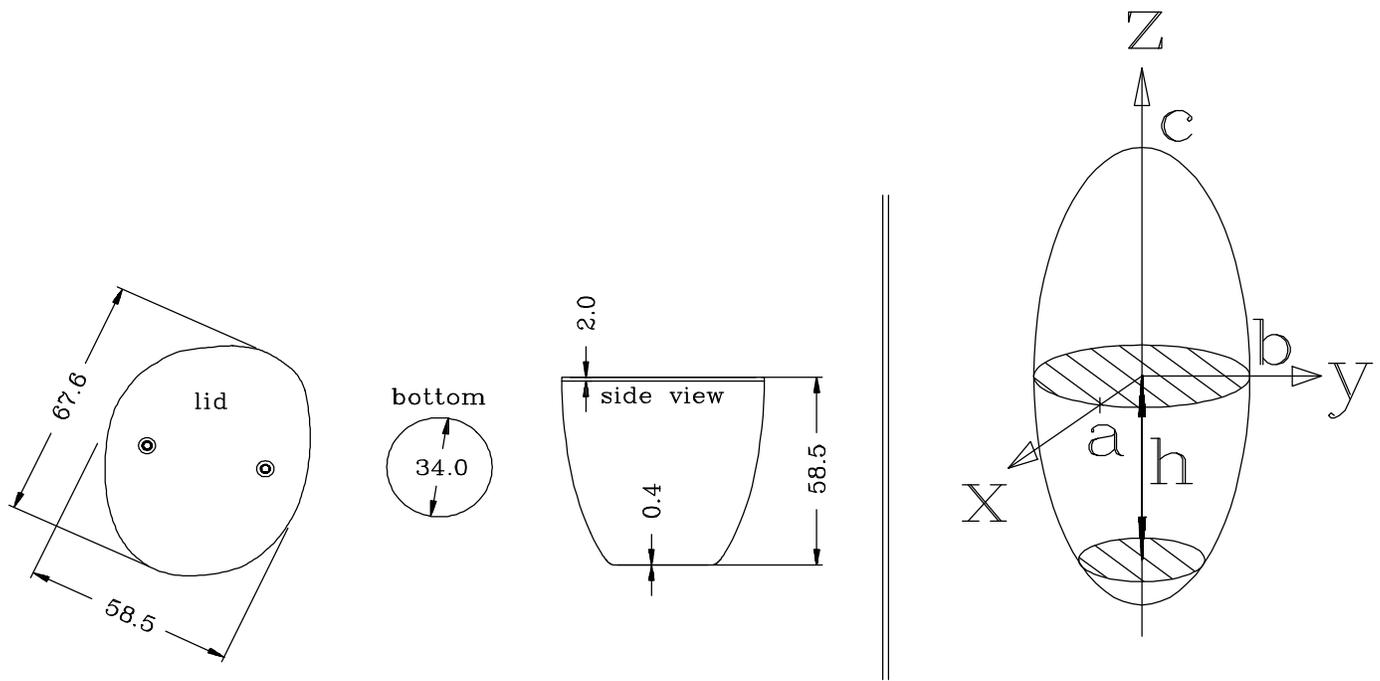
\begin {figure}
\caption{Geometry of the 3D-cup. The investigated microwave resonator
together with its measures (in mm) is given on the left side.
The model of the barrel billiard which was used for several numerical
simulations is shown on the right side. The latter has been constructed
from one half of a three-axial ellipsoid cut with an additional
plane at distance h from the origin.}
\end {figure}

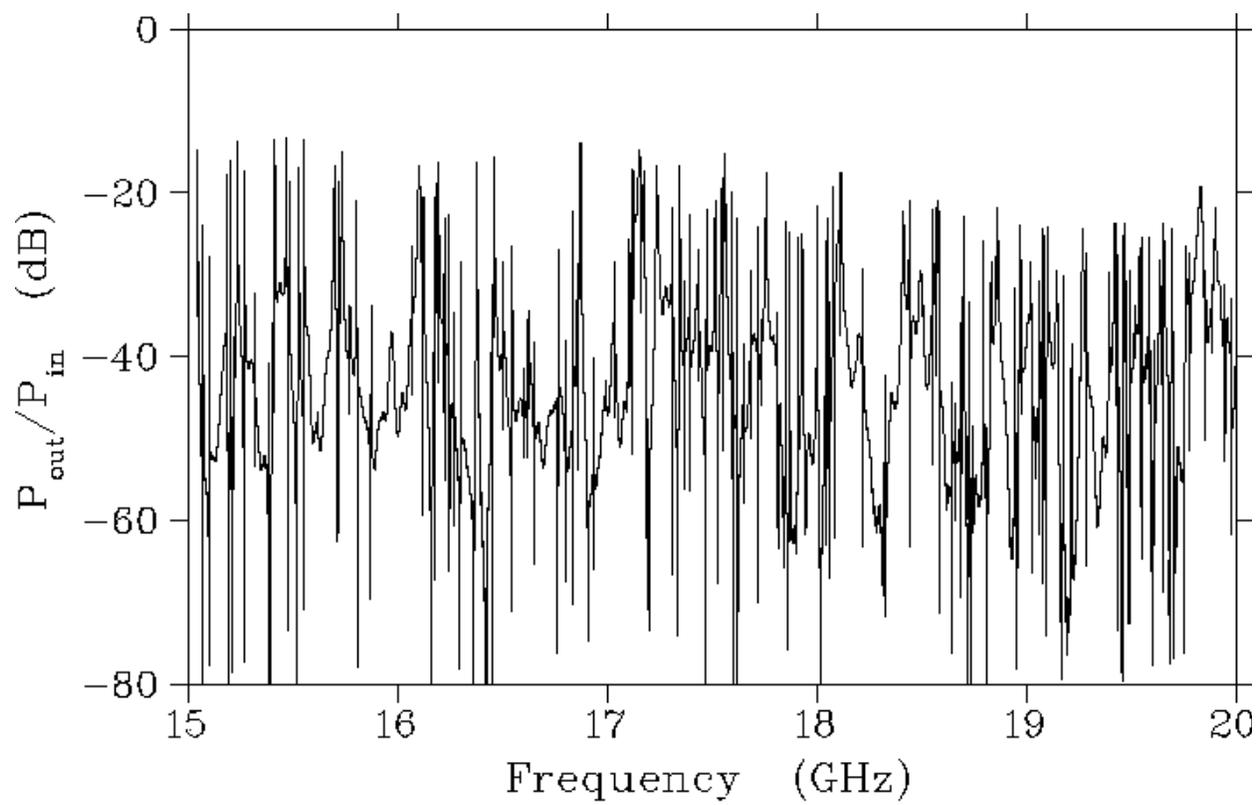
\begin {figure}
\caption{Measured transmission spectrum in a range
between 15 and 20GHz. The ordinate shows the ratio of the output power
relative to the input power on a logarithmic scale. As can be seen from 
the figure the eigenfrequencies appear as sharp peaks with a Q-value
of up to $10^5$ and a signal-to-noise ratio of up to 50dB.}
\end{figure}

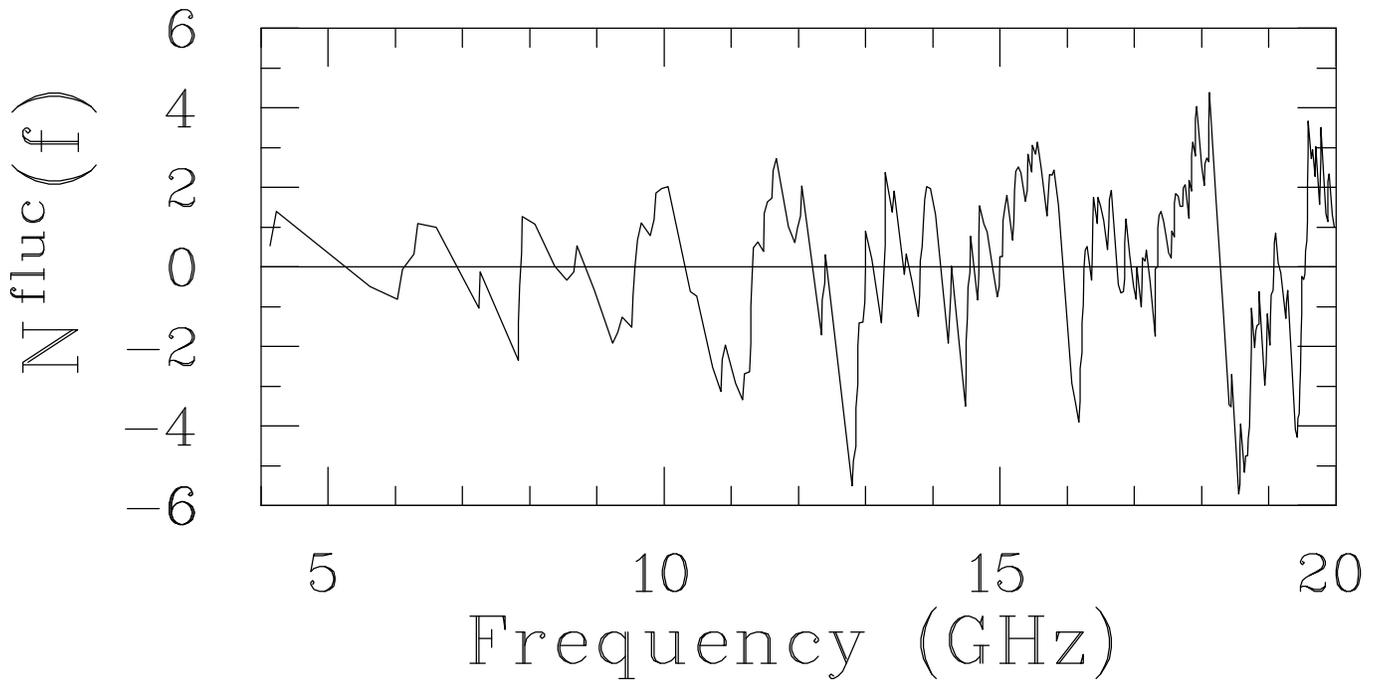
\begin {figure}
\caption{Remaining fluctuating part of the staircase function after the
extraction of the smooth part 
$N^{smooth}(f)=V_1\cdot f^3+V_2\cdot f+V_3$. The
particular form of the oscillations around 
zero indicates that the smooth part was
described consistently.}
\end{figure}

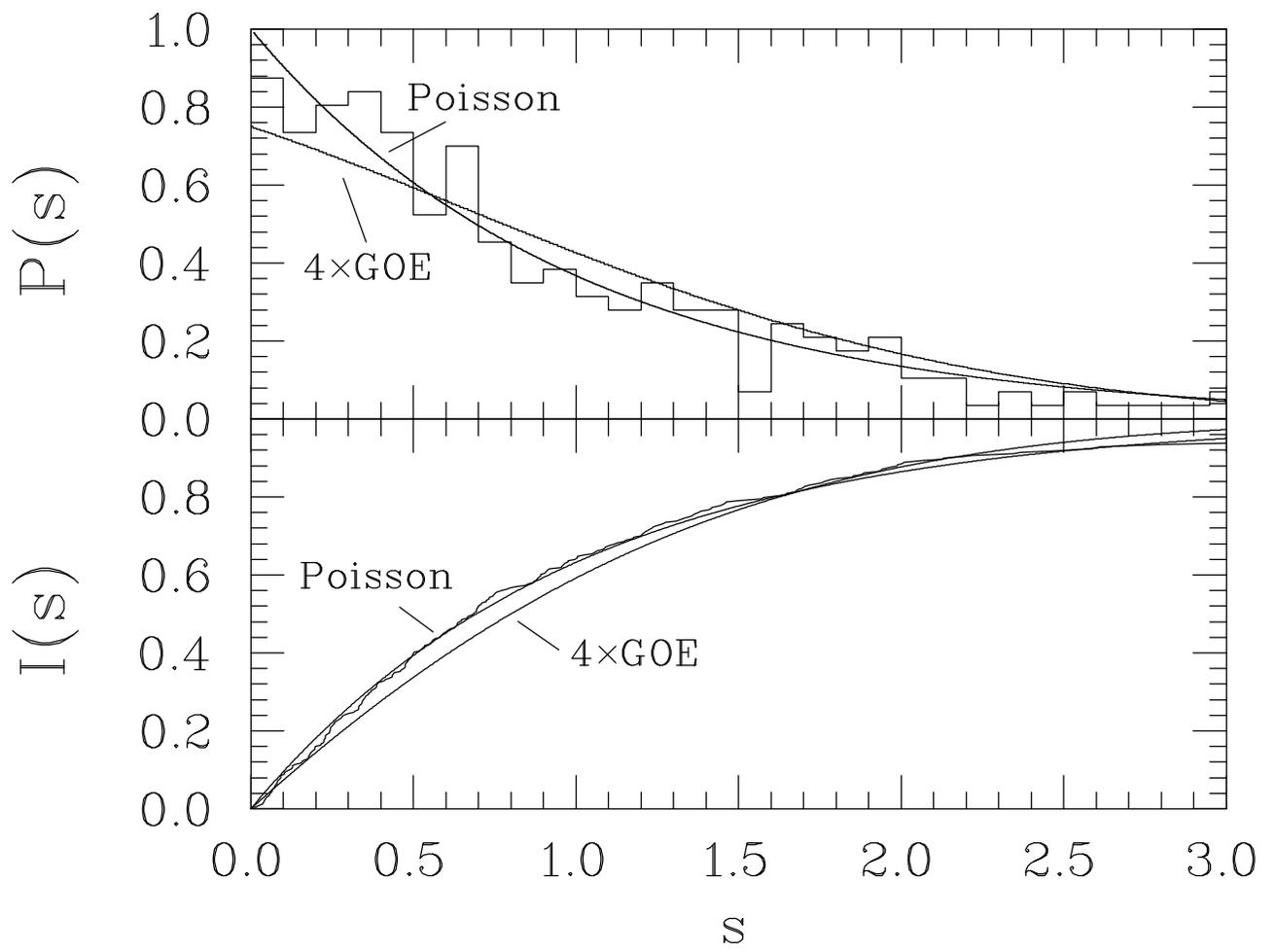
\begin {figure}
\caption{Ordinary (upper part) and cumulative (lower part)
nearest neighbour spacing distribution for the set of 286 measured
eigenfrequencies. Beside the experimental histograms also the
curves for pure Poissonian as well as pure GOE characteristics
are shown.} 
\end{figure}

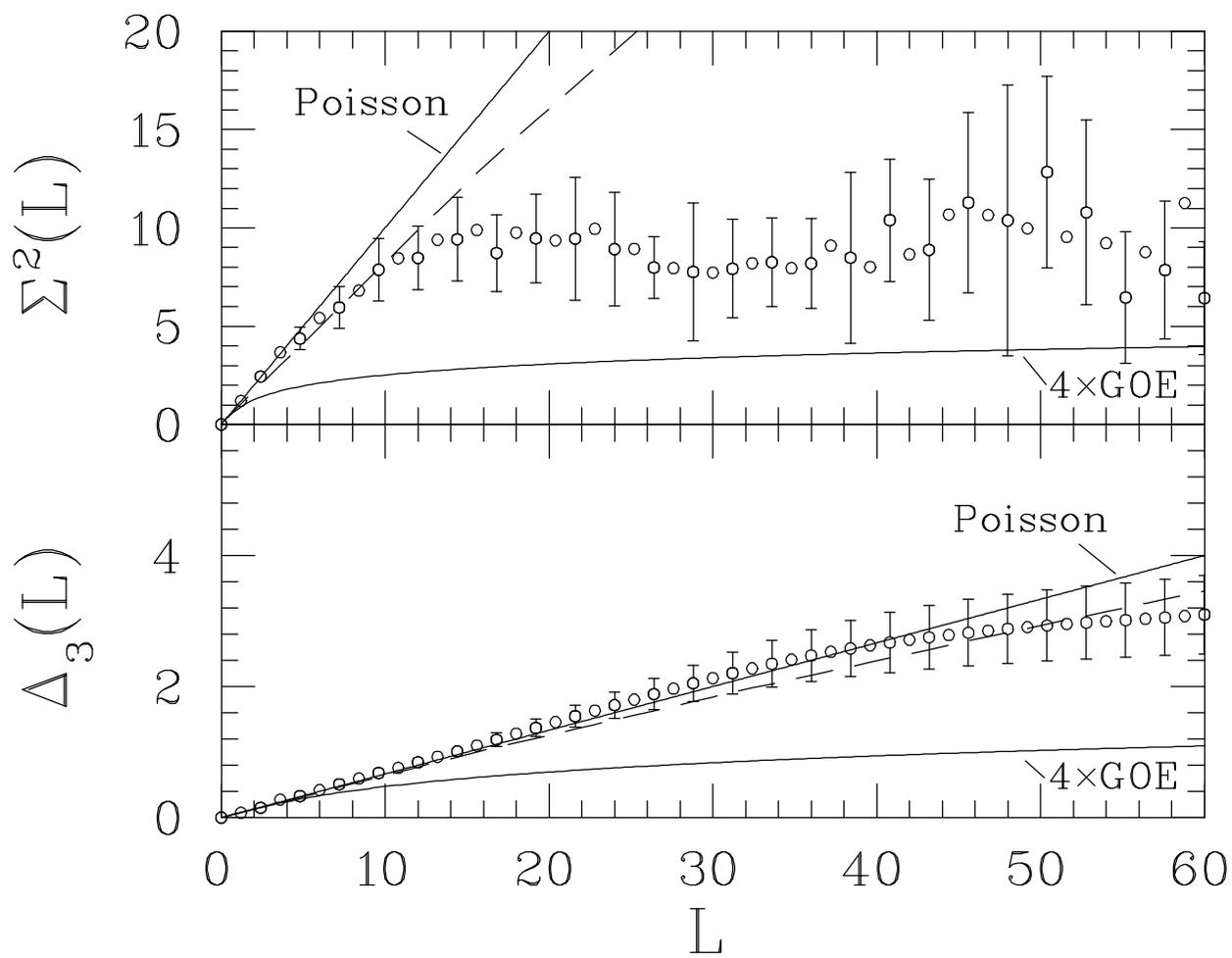
\begin {figure}
\protect\caption{Number variance $\Sigma^2$ and spectral rigidity $\Delta_3$
for the measured spectrum (circles). 
Again the limiting curves for pure Poissonian
and Gaussian distributions are given in the figure. Up to a certain
value $L_{max}$ the $\Sigma^2$-curve is well described by a mixing-parameter
$q\approx0.30\pm^{0.20}_{0.30}$ (dashed curve). 
Above $L_{max}\approx 10$ 
the curve shows a clear deviation which
is theoretically expected because of the finite set of eigenmodes and 
the finite lengths of the shortest periodic orbits. The experimental
spectral rigidity $\Delta_3$ is only
within its error range compatible with the mixing-parameter of $\Sigma^2$
and possesses a four times larger value of $L_{max}$ as expected
\protect\cite {Jost}. For a clear representation only every second
error bar is shown in the figure.}
\end{figure}

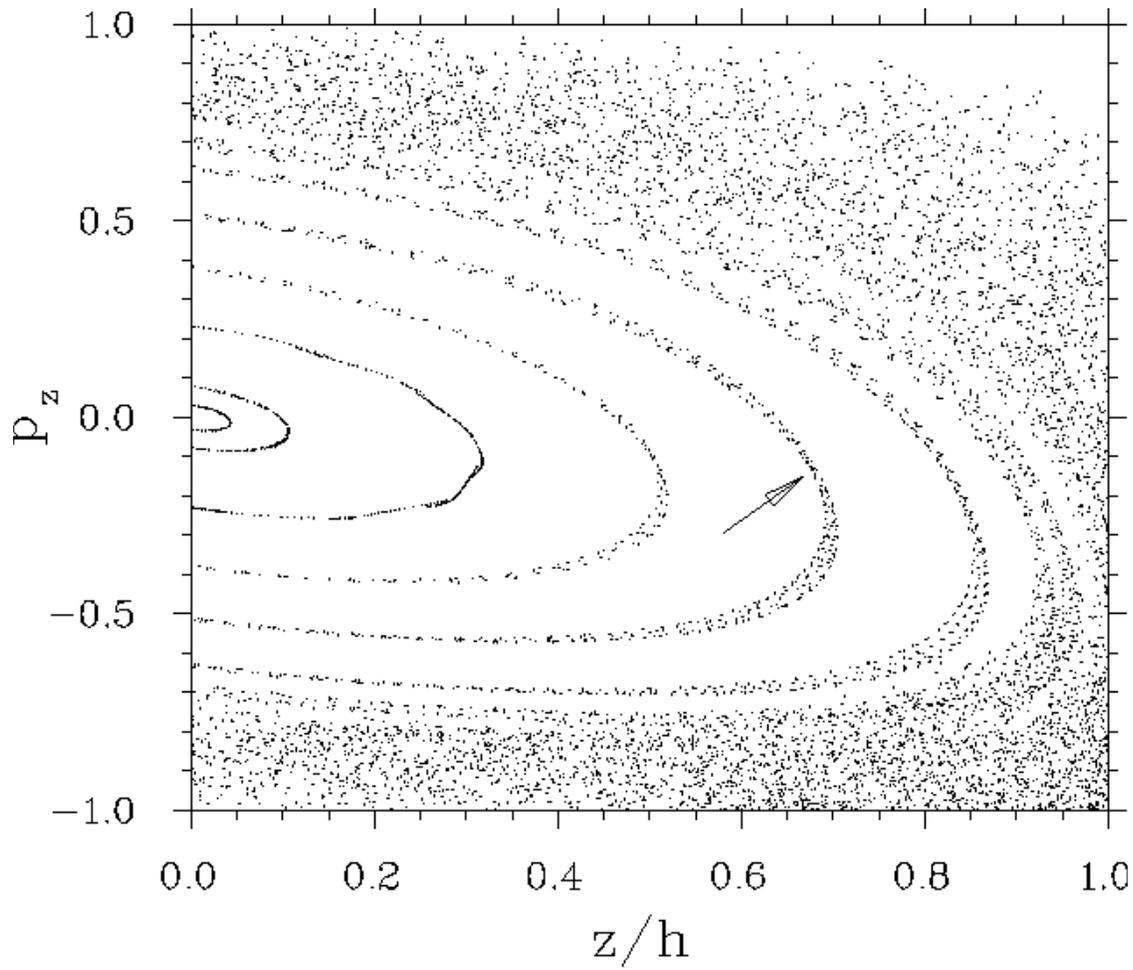
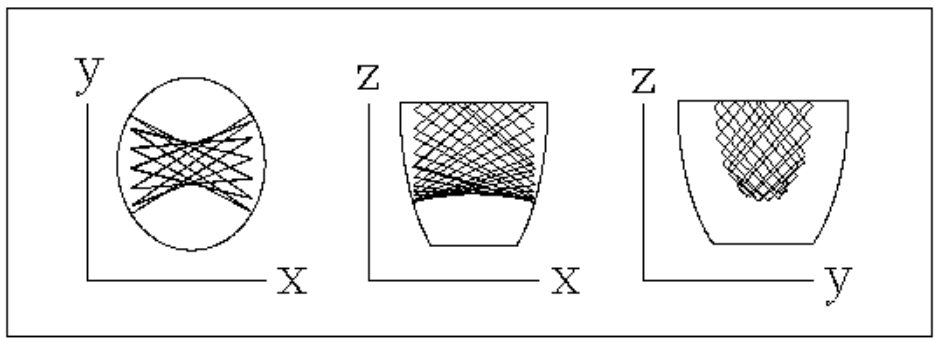
\begin {figure}
\caption{Poincar\'e surface of section for the classical barrel billiard.
The figure shows the resulting patterns for the conjugated variables
($z,p_z$) after 16000 collisions with the boundary. Note that the
momentum $|\vec{p}|$ is normalized to unity. As can be seen 
the phase space is split in regular stripes and a chaotic sea.
The former are produced by a special class of stable orbits one of which
is shown in projections in the lower box of the figure producing one
certain strip pointed by an arrow. These orbits
do not hit the bottom of the cup, thus they only ''see'' the regular
ellipsoid.}
\end{figure}

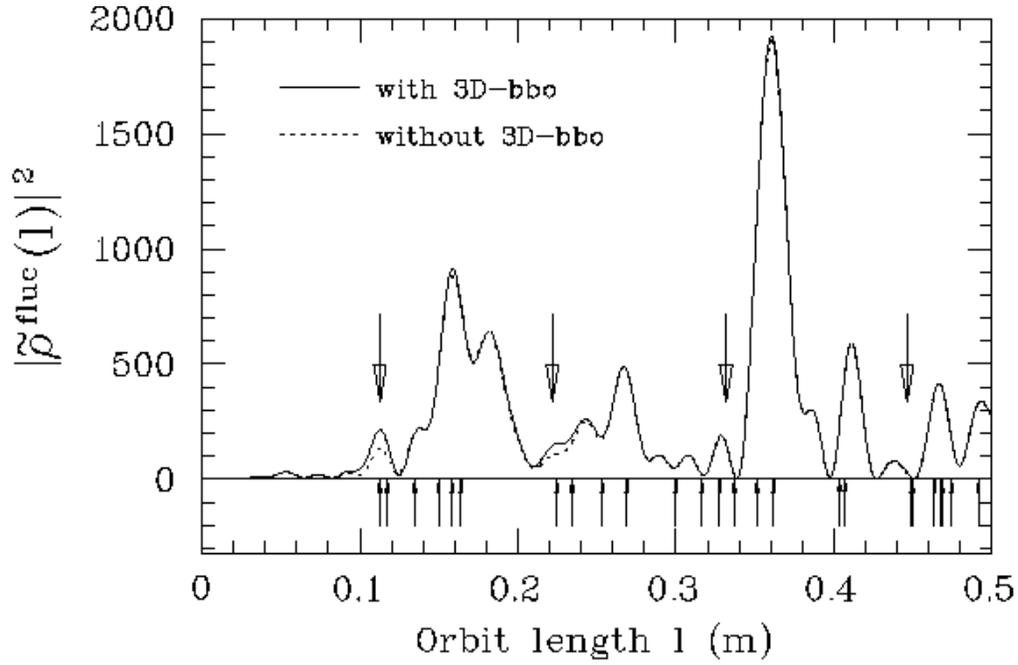
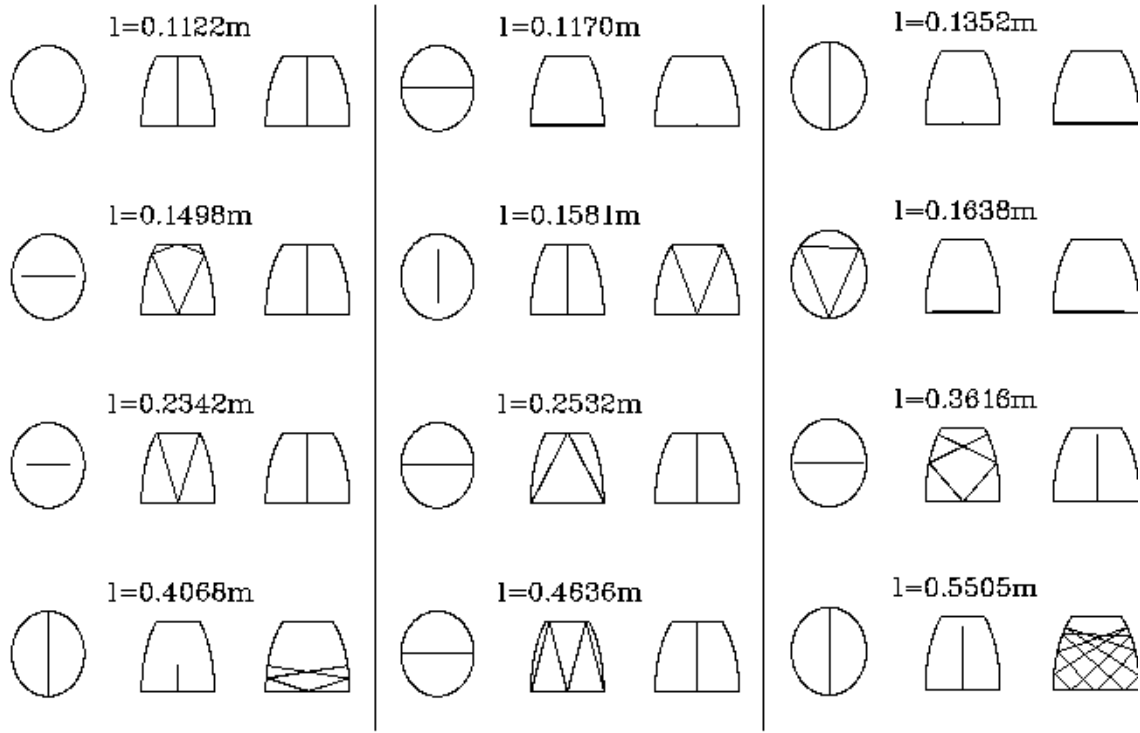
\begin {figure}
\caption{Length spectrum (upper part) from 
the Fourier transformed fluctuating 
part of the eigenmode density of the 3D-cup (solid curve) and some
numerically simulated periodic orbits for the barrel billiard (lower part).
The lengths of these orbits and of their multiples
are indicated by arrows below the abscissa 
for comparison with the experiment. 
In addition the contribution of the
3D-bouncing ball orbit with $l_{bbo}=0.1122$m 
has been extracted from the spectrum (dashed curve),
the larger arrows above the curve 
indicate multiples of $l_{bbo}$ to guide the eye.}
\end{figure}

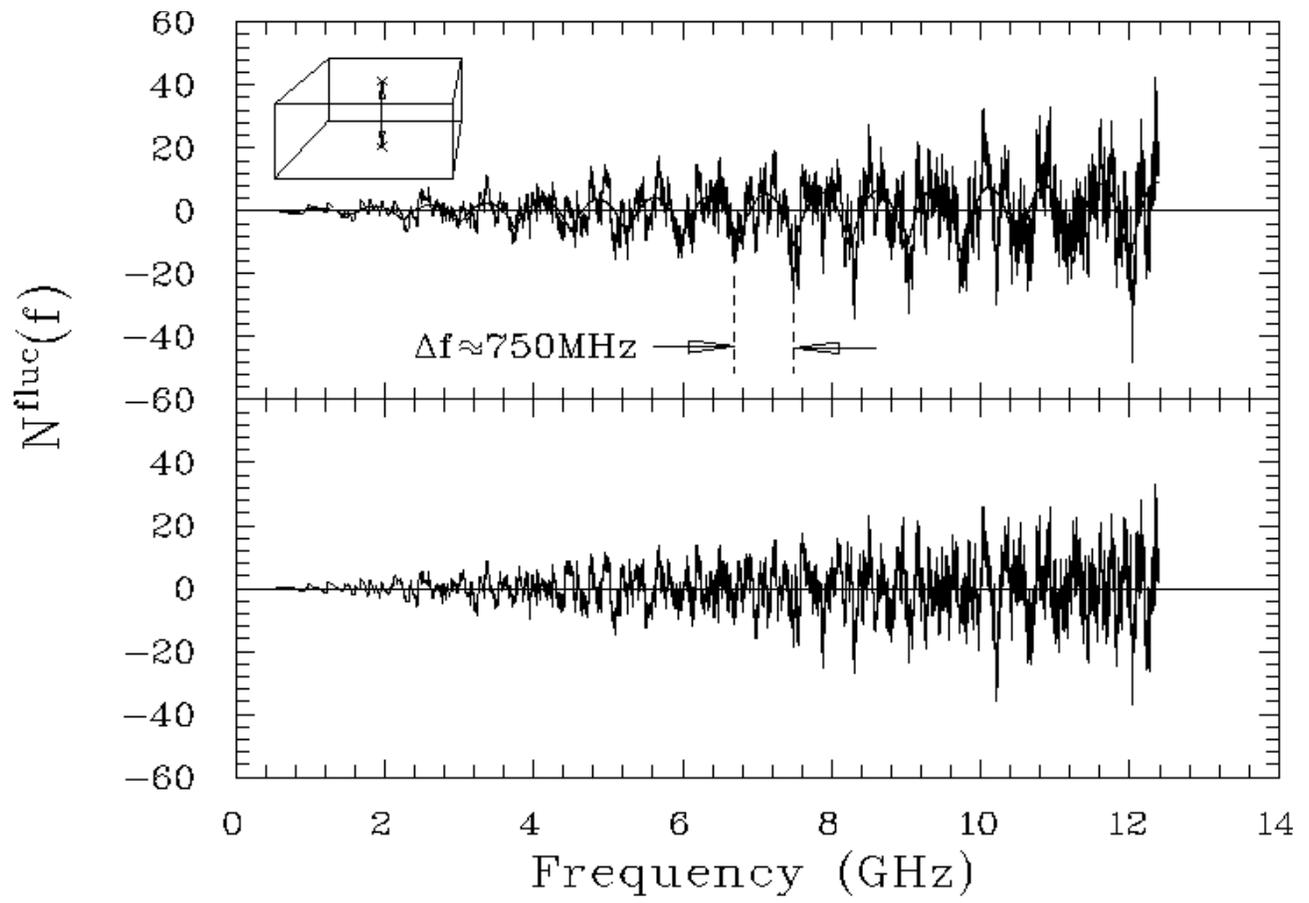
\begin {figure}
\caption{Fluctuating part of the staircase function for a regular
box using the first 20000 eigenfrequencies. In the upper part an
impressive oscillation, with period $\Delta f\approx 750$MHz,
which is due to the contribution of
the shortest 3D-bouncing ball orbit (shown in the insert) can be observed.
The theoretical investigation, Eq.(\ref{3DBB2}), 
resulted in the smooth curve also
superimposed in the upper part of the figure. After the extraction of
this bouncing ball contribution the oscillations around zero revealed
a lower magnitude and the corresponding period $\Delta f$ vanished
from the fluctuations (lower part).}
\end{figure}

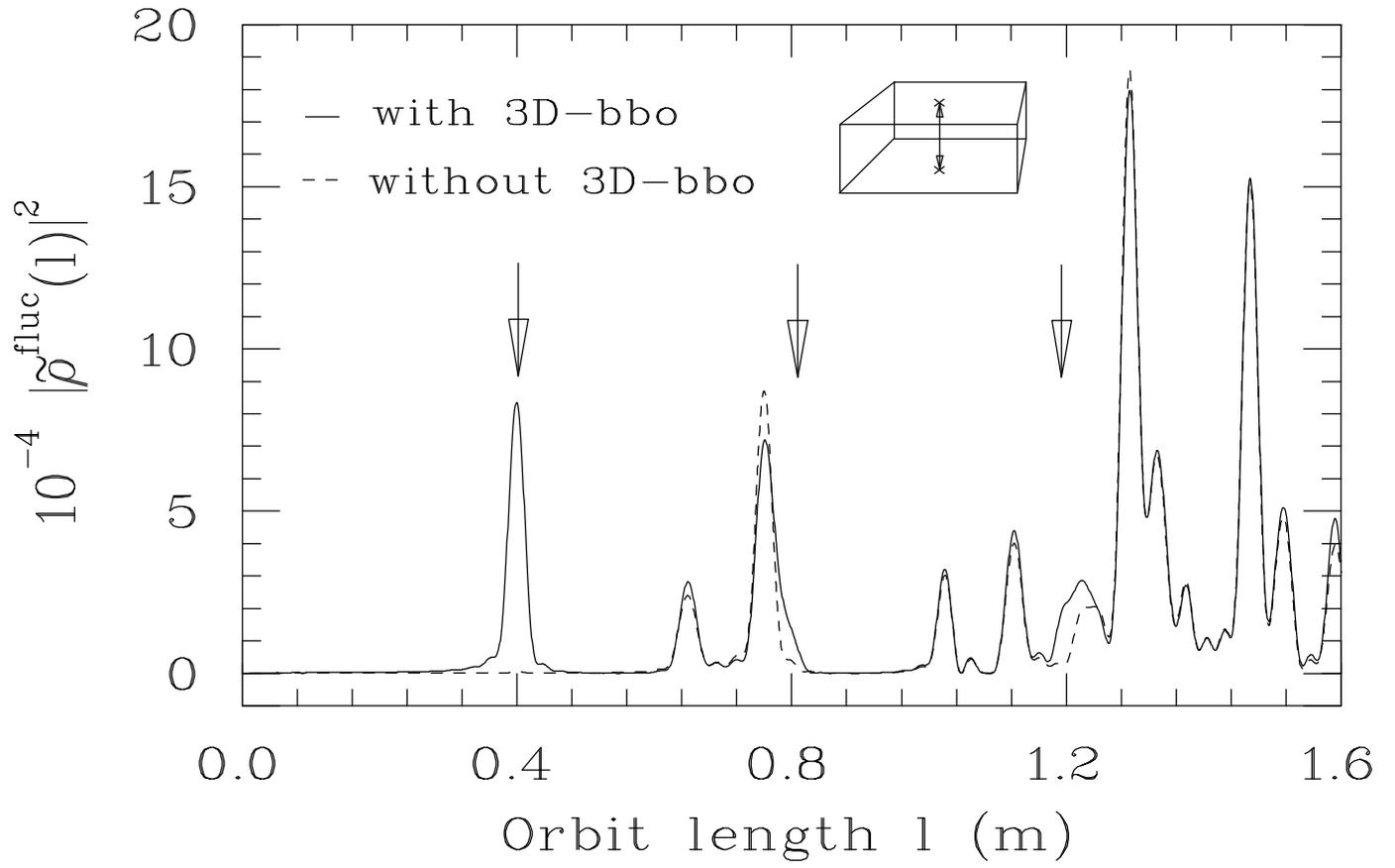
\begin {figure}
\caption{Length spectrum for the regular box in a range between 0 and 1.6m
(solid curve).
In this presentation the discussed bouncing ball orbit leads to an impressive
peak at $l_{bbo}=0.4$m. The positions of its multiples are 
denoted by arrows in the figure. Again the contribution of this orbit
has been extracted and now the remaining spectrum (dashed curve)
can be seen to
be free of any remnant.}
\end{figure}

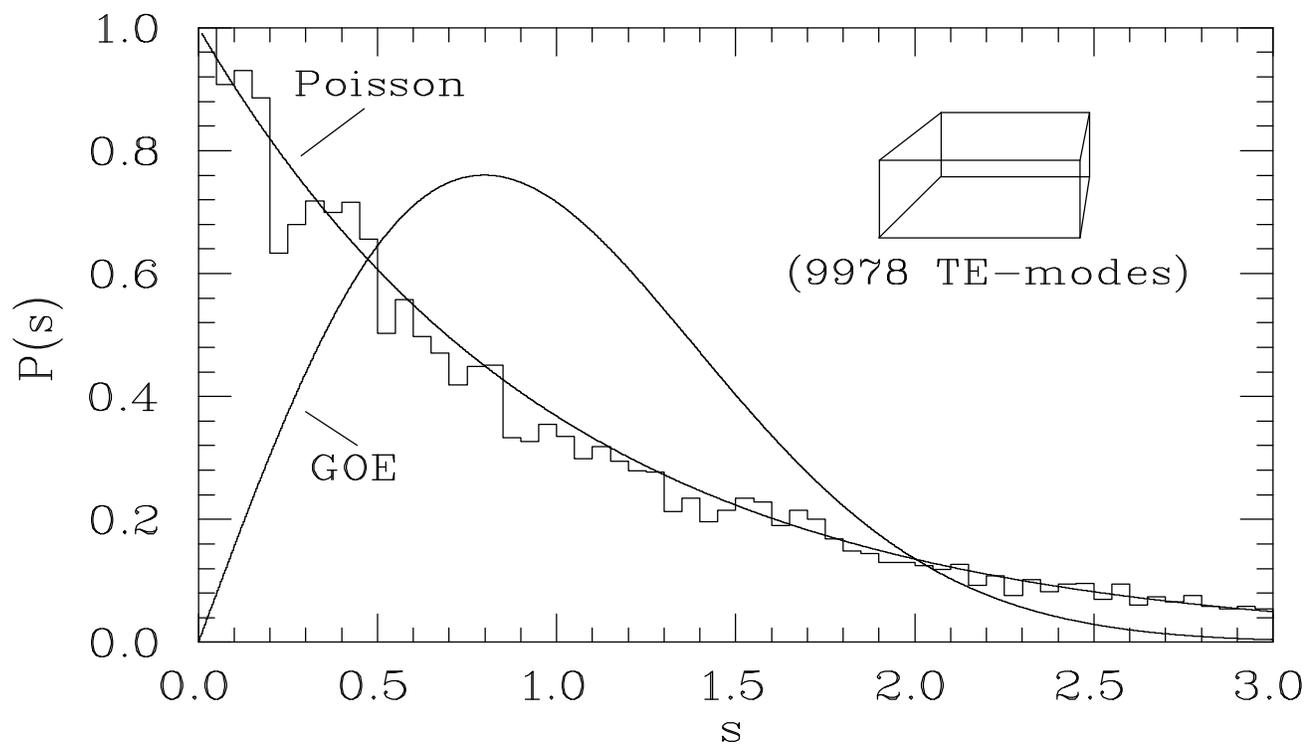
\begin {figure}
\caption{Nearest neighbour spacing distribution for the regular box. Because
of the systematic degeneration of almost 
all TE- and TM-modes, the calculation
was restricted to one polarization (9978 TE-modes). Due to the fact that
the system is desymmetrized per definition
the limiting curve for the chaotic case
is given by a single GOE distribution. As expected for the regular
box the system shows pure Poissonian characteristics.}
\end{figure} }

\end{multicols}

\end{document}